\documentclass[%
 aps,
 pra,
 amsmath,amssymb,
 preprint,%
]{revtex4-1}

\usepackage{color}
\usepackage{graphicx}
\usepackage{dcolumn}
\usepackage{bm}

\draft 

\begin{document}


\title{{A scheme for realizing continuously tunable spectrum in the visible light region based on monatomic carbon chains}} 



\author{Chen Ming}
\email[]{These authors contribute equally to this work}
\affiliation{Institute of Modern Physics, Fudan University, Shanghai 200433, China}
\affiliation{Applied Ion Beam Physics Laboratory, Key Laboratory of the Ministry of Education, Fudan University, Shanghai 200433, China}

\author{Fan-Xin Meng}
\email[]{These authors contribute equally to this work}
\affiliation{Institute of Modern Physics, Fudan University, Shanghai 200433, China}
\affiliation{Applied Ion Beam Physics Laboratory, Key Laboratory of the Ministry of Education, Fudan University, Shanghai 200433, China}

\author{Xi Chen}
\affiliation{Institute of Modern Physics, Fudan University, Shanghai 200433, China}
\affiliation{Applied Ion Beam Physics Laboratory, Key Laboratory of the Ministry of Education, Fudan University, Shanghai 200433, China}

\author{Jun Zhuang}
\affiliation{Department of Optical Science and Engineering, Fudan University, Shanghai 200433, China}

\author{Xi-Jing Ning}
\email[]{Electronic address: xjning@fudan.edu.cn}
\affiliation{Institute of Modern Physics, Fudan University, Shanghai 200433, China}
\affiliation{Applied Ion Beam Physics Laboratory, Key Laboratory of the Ministry of Education, Fudan University, Shanghai 200433, China}


\date{\today}

\begin{abstract}
We propose a scheme for realizing the continuously tunable spectrum based on monatomic carbon chains. By hybrid density functional calculations, we first show that the direct band gap of monatomic carbon chains change continuously from 1.58 to 3.8 eV as strain is applied from -5 to 10\% to the chain, with separated Van Hove singularity peaks enhanced. To realize this tunability, a realistic stretching device is proposed by contacting the chain with graphene sheets, which can apply up to 9\% elongation to the chain, yielding tunable light-emitting wavelengths from 345 to 561 nm.
\end{abstract}


\maketitle 

\section{Introduction}
The wavelength tunability and electro-optical conversion efficiency are the most important parameters for opto-electronic devices such as lasers and light emitting diodes (LEDs). During the last decade, the development of semiconductor lasers has greatly improved the electro-optical conversion efficiency up to more than 10\% and even reached 50\%\cite{1}; meanwhile, the range of wavelength has been largely extended to cover from 350 to nearly 2000 nm with different working mediums and partially tunable laser around certain wavelengths were also realized\cite{1,2}. Nevertheless, the output wavelength is still far from continuously tunable in the spectra range as intervals of tens of nanometers exist between different bands\cite{2}, which limits their applications to accurate spectroscopy and high sensitive detectors\cite{2,3}.

Recently, carbon nanotubes(CNTs) have been suggested as promising working mediums for nanoscale tunable lasers and LEDs. Because CNTs own direct band gaps and quasi one-dimensional(1D) structures, which lead to the diverging density of states(DOS) near the band edges, known as Van Hove singularities(VHSs)\cite{4}, and were expected to produce strong light emissions with very narrow bands and wide wavelength tunability, which have been recently confirmed by intensive experiments\cite{5,6,7,8,9,10,11}. Compared with CNTs, monatomic carbon chains(MCCs) can be more fascinating candidates in principle, because they have true 1D structures, and are expected to exhibit uniquely featured VHS peaks along with better tunability. The structure of infinite long MCCs feature the bond length alternation(BLA) known as Peierls distortion(PD)[Fig. 1(a)], which opens a direct band gap in MCCs. This gap was predicted to be about 0.3 eV and can be continuously tuned to 1.5 eV as the chain is elongated by up to 10\% according to density functional theory(DFT) calculations with General Gradient Approximation(GGA)used\cite{12,13}. Based on this result, a tunable laser based on MCCs has been theoretically proposed\cite{13}.

Recent progresses in synthesis and characterization of MCCs provide valuable data to test the above theoretical predictions. Stable and rigid MCCs with a length of 2.1 nm(about 16 atoms) were observed inside a transmission electron microscope\cite{14}, and a series of polyyne, end capped monoatomic carbon chain with up to 44 atoms have been fabricated and characterized\cite{15,16,17}. By measuring the dependence of properties of these polyynes on the chain length, it is predicted that chains with more than 48 atoms should exhibit the converged properties for the infinite length\cite{17}, and the converged values for BLA $\delta$ ($\delta = L_1-L_2$, $L_1$ and $L_2$ are the long and short C-C bond length) and the band gap $E_g$  are 0.135 \AA\cite{16} and 2.2 eV\cite{15}(or 2.56 eV\cite{17}). However, the predictions of the GGA-DFT are only 0.04 \AA\ and 0.3 eV\cite{12,13}, which are far smaller than the experimental estimations. We note that such severe underestimations have also been found in the conjugated polymers as localized density approximation(LDA) and GGA-DFT are used. A typical example is \textit{trans}-polyacetylene(\textit{t}-PA), where the calculated $\delta$ and $E_g$ by GGA-DFT are 70\% and 90\% smaller than the well measured values\cite{18}. Thus, to make device designs for MCCs, their structural and electronic properties along with the tunability should be reevaluated by a more reasonable method.

Considering that the hybrid density functional can well predict the properties of conjugated polymers\cite{18}, we employed the Heyd-Scuseria-Ernzerhof hybrid density functional(HSE06) \cite{19,20} in this work to investigate the structural and electronic properties of the infinite long MCC. As the tunability of electronic structure is the focus of the present work, we think the infinite long model will reflect the main features of finite long chains; On a realistic aspect, chains with more than 48 atoms are estimated to behave as the infinite long model, which are not very far from the currently fabricated chains\cite{20}. The empirical parameter in HSE06 was firstly optimized to fit the experimentally obtained $\delta$ and $E_g$ of \textsl{t}-PA. Because the quasi-chain structure of \textsl{t}-PA is very similar to that of MCCs, the optimized hybrid functional is expected to reproduce the experimental values of $\delta$ and $E_g$ for MCCs. Then this functional was used to evaluate the properties of stretched MCCs.

\section{Calculation methods}
The HSE06 hybrid functional in our DFT calculation reads \cite{19,20}:
\begin{equation} \label{1}
E_{xc}^{HSE}=xE_x^{HF,SR} + (1-x)E_x^{PBE,SR} + E_x^{PBE,LR} + E_c^{PBE}
\end{equation}
where the short ranged Hatree-Fock(HF) exchange energy $E_x^{HF,SR}$ is combined with the short ranged Perdew-Burke-Ernzerhof (PBE)-GGA exchange energy $E_x^{PBE,SR}$ by a parameter $x$, plus the long ranged PBE-GGA exchange energy $E_x^{PBE,LR}$ and the PBE-GGA correlation energy $E_c^{PBE}$. The projector augmented wave(PAW) potentials and plane wave basis sets were employed in the calculations implemented by Vienna Ab-initio Simulation Package\cite{21}. Considering the BLA featured structures, the unit cell of MCCs is set to contain two carbon atoms [Fig. 1(a)], and the \textsl{t}-PA cell contains two carbon and two hydrogen atoms [Fig. 1(b)]. For both structure relaxation and band structure calculations, a $40\times1\times1$ k point grid was found to reach the convergence and was used for all these calculations. The maximum Hellmann-Feynman forces acting on each atom is less than 0.01 eV/\AA\ for structure relaxation.

\section{Results and discussion}
\subsection{The structural and electronic properties of MCCs}
To optimize the parameter $x$ in Eq. (1), the structure and the band gap of \textsl{t}-PA were calculated with $x$ changing from 0 to 0.5. As shown in Fig. 2, both $\delta$ and $E_g$ increase monotonically with $x$ increasing. The optimized value $x_0=0.42$ was obtained by fitting the experimentally value of $E_g$ 1.48 eV\cite{22}, and then the corresponding $\delta$ obtained  in Fig. 2(b) is 0.069 \AA, very close to the experimental values 0.08 \AA\cite{18,23}, and the C-C bond length 1.419 and 1.351 \AA\ are also close to the experimental estimations 1.44$\sim$1.46 and 1.36$\sim$1.39 \AA\cite{18,23}. It is notable that the calculated $\delta$ and $E_g$ by the standard PBE-GGA DFT ($x=0$) are 0.013 \AA\ and 0.12 eV, one order of magnitude smaller than that experimentally measured.

We then applied different $x$ (0$\sim$0.5) to the calculations of MCCs, and found a similar increasing of $\delta$ and $E_g$ with $x$ to that of \textsl{t}-PA [Fig. 2]. As $x_0$ optimized from \textsl{t}-PA was used, the bond length of MCCs were calculated as 1.332 and 1.219 \AA, and the corresponding $\delta$ and $E_g$ are 0.113 \AA\ and 2.21 eV. These value matches quite well with the experimental extrapolated $\delta$ and $E_g$ for the infinite long MCC, 0.135 \AA\cite{16} and 2.2 eV\cite{15} (or recently corrected to 2.56 eV\cite{17}), suggesting that the HSE06 hybrid functional with $x_0$ is capable of describing the main structural and electronic properties of MCCs at equilibrium. Moreover, as this parameter accurately describes two different linear carbon systems(\textsl{t}-PA and MCCs) at equilibrium, we expect it can give reasonable predictions as MCCs are stretched (in calculation below). Compared with the $E_g$ of \textsl{t}-PA (1.48 eV), the MCC own a much larger band gap, implying that it should own much stronger PD than \textsl{t}-PA. To show this fact exactly, the reduced BLA $\delta'$ ($\delta'=\delta/(L_1+L_2)$) for these two structures were evaluated, and the $\delta'$ of the MCC (4.4\%)is indeed larger than that 2.5\% of \textsl{t}-PA.

\begin{figure}
\includegraphics[width=5.8cm]{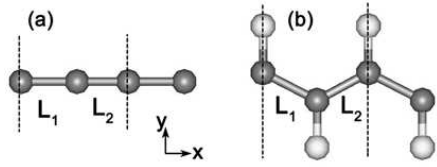}%
\caption{Unit cells used in the calculations of MCCs, with 2 carbon atoms(a) and of \textsl{t}-PA, with 2 carbon and 2 hydrogen atoms(b). The long and short C-C bond length are denoted by $L_1$ and $L_2$, and BLA $\delta=L_1-L_2$.}%
\end{figure}

\begin{figure}
\includegraphics[width=8.5cm]{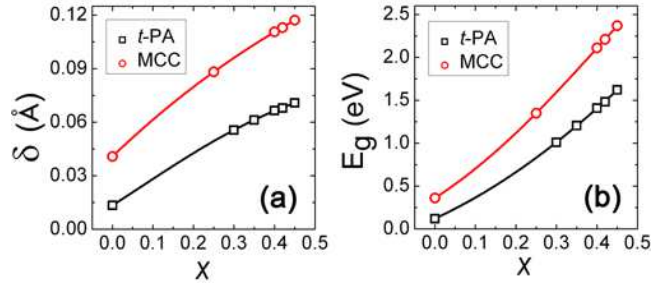}%
\caption{(Color online) Changes of BLA $\delta$ (a) and the band gap $E_g$ (b) with the empirical parameter $x$ changing from 0 to 0.5 in \textsl{t}-PA (square) and MCCs (circle).}
\end{figure}

 With the optimized functional applied, the calculated band structure of the MCC features a direct band gap [Fig. 3(a)]. The valence and the conduction bands are two-fold degenerated, which are formed by the two unbonded $p$ orbitals of carbon atoms in this $sp$ bonded chain. This feature should account the stronger PD in the MCC, as there is no degeneracy in the bands of \textsl{t}-PA. As seen in Fig. 3(c), the DOS of the MCC features sharp peaks located at the band edges (exactly at the VHSs),whcih are resulted from the 1D chain structures. The existence of these peaks implies that MCCs can produce strong light emissions with narrow bands because the quantum transition rate is proportional to the number of the electronic states. This implication is supported by CNTs, as they own similar VHS-peak structures [Fig. 3(c)] and the strong light emissions with quite narrow bands have indeed been observed by extensive experiments\cite{5,6,7,8,9,10,11}. Notably, the VHS peaks in the MCC are much more separated from each other compared with those of CNTs, as the distance between the first and second VHS peaks above the Fermi level is about 3 eV [Fig. 3(b)], which is much larger than that of CNTs, such as less than 0.5 eV for (10,0) tubes[Fig. 3(c)], indicating that much stronger and purer spectral light emissions would be obtained in MCCs. Moreover, as there are fewer phonon modes in MCCs than in CNTs, the non-radiative channels should be fewer in MCCs and the electro-optical efficiency could be high.

\begin{figure}
\includegraphics[width=8.5cm]{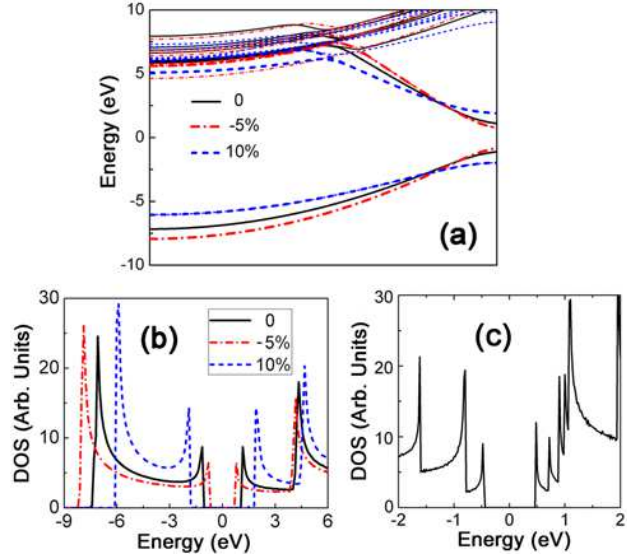}%
\caption{(Color online) Calculated (a) band structures and (b) DOS of MCCs with different strains applied by the optimized HSE06, where the solid, dot-dashed and dashed lines corresponds to 0, -5\% and 10\% strain, respectively. (c) DOS of the (10,0) armchair CNTs. The center of the band gap is set as zero point.}
\end{figure}

\subsection{Tuning the properties of MCCs by stretching them}
To study the property of the stretched MCCs, we first investigated their mechanical properties using the optimized HSE06 hybrid functional $x_0$ (0.42), and the PBE-GGA DFT was also used for comparison. The strain-stress curve [Fig. 4(a)] suggests that these two methods yield very similar mechanical behaviors for the MCC, particularly in the elastic area (within -5\%$\sim$10\% elongation as denoted in Fig. 4(a)), and the calculated 1D elastic modulus $C^{1D}$ is 81 eV/\AA\ for GGA and 90 eV/\AA\ for HSE06 (The 1D elastic modulus is defined as $C^{1D} = \partial F/\partial\varepsilon$ and was extracted from the linear region of Fig. 4(a), where $F$ is the applied force and $\varepsilon$ is the corresponding strain). The ultimate strain yielded by HSE06 is 20.8\% with a breaking force of 12.2 nN, which is also similar to that obtained by GGA-DFT 17.6\%, 11.5 nN. This result suggests that the dimerization of carbon atoms has little effect on the mechanical property of MCCs, and the GGA-DFT calculations are sufficient to make a reasonable description. We note that our calculated breaking force matches perfectly the experimental result of 11.2 nN\cite{24}, and the ultimate strain is also consistent with other calculations\cite{25,26}.

\begin{figure}
\includegraphics[width=6.5cm]{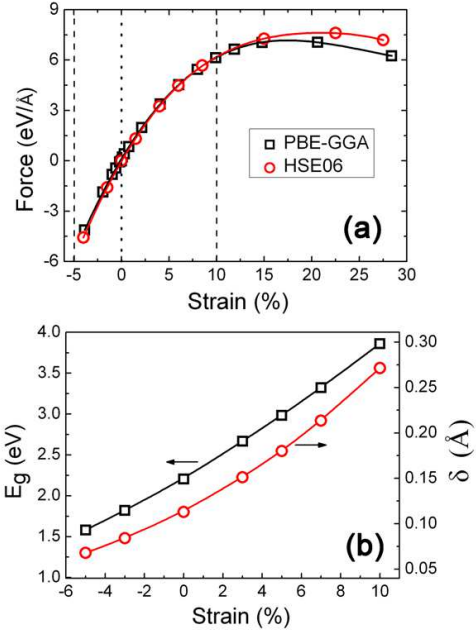}%
\caption{(Color online) (a) Calculated strain-stress curves of MCCs by PBE-GGA DFT (square) and the optimized HSE06 (circle), the area between the two dashed lines are the elastic strain area focused. (b) Changes of and of MCCs with changing strains calculated by the optimized HSE06.}%
\end{figure}

Within the elastic area, the structures and the electronic properties of the stretched MCCs were calculated with the optimized hybrid functional. As the strain is continuously increased from -5\% to 10\% [Fig. 4(b)], $\delta$ and $E_g$ increase from 0.068 to 0.27 \AA\ and from 1.58 to 3.86 eV, respectively, and the direct gap feature keeps within all this elastic area. Since the bottom (top) of the conduction (valence) bands is pushed up (down) as the chain is stretched, as shown in Fig. 3(a), the slopes of these bands gets much gentler near the Fermi level, which makes the VHS peaks become much sharper [Fig. 3(b)]. This indicates stronger and narrower-band light emissions can be expected via the elongated MCCs. The reduced BLA increases from 2.8\% to 9.7\% as the chain is continuously stretched from -5\% to 10\% elongation, indicating that PD is enhanced by increasing elastic strain, and this explains the band gap widening. Compared with bulk semiconductors, where the band gap tunability is quite limited because of poor ductility ($<$1\%), the tunability of the band gaps in MCCs (1.58$\sim$3.86 eV) is much significant, making them potentially good mediums for wavelength tunable light emitting devices.

\begin{figure}
\includegraphics[width=7.5cm]{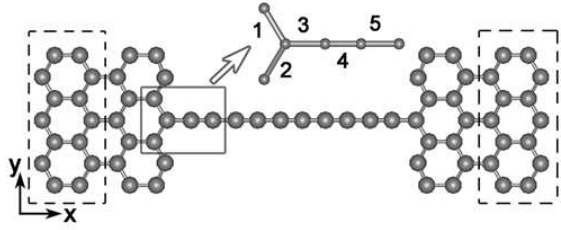}%
\caption{A proposed stretching device contacting a 10-atom long MCC with two large grahene pads. The inset zooms in on the contacting structure, where the numbers denote different C-C bonds. Bond 1 and 2 denote the bonds in graphene, bond 3 is the contacting bond between graphene and the MCC, bond 4 and 5 are bonds in MCCs, and the corresponding bond length is 1.434, 1.434, 1.381, 1.248 and 1.318\AA\ respectively.}
\end{figure}

\subsection{A realistic device proposed to stretch MCCs}
To realize the above gap tuning process of MCCs, we considered a realistic stretching device by contacting a 10 atom-long chain with two large graphene pads [Fig. 5], which has indeed been experimentally realized\cite{14}, and by stretching the two graphene pads, the chain can be continuously elongated. To find the maximum strain that this device can apply on the chain, we performed a GGA-DFT calculation using the model in Fig. 5. Periodical boundary conditions were applied on $x-y$ plane to simulate that the chain is contacted with two infinite large graphene pads. To stretch the device, the box size is gradually enlarged in the $x$ direction up to 15\% elongation, and at each elongation point the whole device structure was optimized. When the device was initially stretched, both the graphene pads and the chain are elongated. But as stretching continues, graphene becomes hard to be elongated, and the strain is mainly applied on the chain and the contact between the chain and the graphene pads (bond 3 denoted in Fig. 5). We found that more than 9\% elongation will break this device with a breaking force of 5.8 nN, and the breaking point is at the contacting point. From a bond length analysis (seeing the inset of Fig. 5), we can see that the contact bond (bond 3) is indeed the weakest point of the device. At the ultimate strain of the device, the chain is found to be elongated by 9.1\%. It is notable that although the value 9.1\% is obtained from a 10 atom-long chain, it is a reasonable estimation for even longer chains. Because calculations have shown that both the breaking force and ultimate strain barely change with the chain length increasing\cite{26}, and our calculations indicates the contact bond length is also insensitive to the chain length. Corresponding to 0$\sim$9\% strain, the gap of MCCs can be tuned from 2.21 to 3.59 eV in this device.

For applications, the graphene pads at the two sides are natural electrodes, so the system above can be prototypes of wavelength tunable opto-electronic devices. And the working wavelength can be continuously tuned by stretching the large graphene pads with controlled force. This proposed device may be realized under current experimental conditions as the graphene-MCC contacting structure has been fabricated, and the stretching process is also experimentally possible, but a precise force control is needed. From this device, we showed recently that via doping the MCC, it can work as a lasers by applying bias on the graphene electrodes\cite{13}. Considering newly presented results, the wavelength of this laser may be continuously tuned from 345 to 561 nm.

\section{Conclusion}
In summary, we calculated the structural and electronic properties of the infinite long MCCs by the hybrid functional HSE06 with the optimized empirical parameter $x_0$ from \textsl{t}-PA. The results obtained at the equilibrium chain length agrees well with the experimental measurements. By applying elastic strains from -5\% to 10\% to the chain, its band gap can be changed from 1.58 to 3.8 eV, with the VHS peaks enhanced gradually. To realize this tunability, we considered a realistic graphene-MCC device, and found up to 9\% strain on MCCs can be applied through this device. These results imply the potential applications of MCCs for future nanoscale lasers and other opto-electronic devices with tunable wavelength in the visible region.


%
%

%

\begin{acknowledgments}
The authors are very grateful to acknowledge Professor Qike Zheng for helpful discussions. This work was supported by the National Natural Science Foundation of China under Grant No. 11274073, 11074042 and 51071048, Shanghai Leading Academic Discipline Project (Project No. B107) and Key Discipline Innovative Training Program of Fudan University.
\end{acknowledgments}


\begin{thebibliography}{26}%
\makeatletter
\providecommand \@ifxundefined [1]{%
 \@ifx{#1\undefined}
}%
\providecommand \@ifnum [1]{%
 \ifnum #1\expandafter \@firstoftwo
 \else \expandafter \@secondoftwo
 \fi
}%
\providecommand \@ifx [1]{%
 \ifx #1\expandafter \@firstoftwo
 \else \expandafter \@secondoftwo
 \fi
}%
\providecommand \natexlab [1]{#1}%
\providecommand \enquote  [1]{``#1''}%
\providecommand \bibnamefont  [1]{#1}%
\providecommand \bibfnamefont [1]{#1}%
\providecommand \citenamefont [1]{#1}%
\providecommand \href@noop [0]{\@secondoftwo}%
\providecommand \href [0]{\begingroup \@sanitize@url \@href}%
\providecommand \@href[1]{\@@startlink{#1}\@@href}%
\providecommand \@@href[1]{\endgroup#1\@@endlink}%
\providecommand \@sanitize@url [0]{\catcode `\\12\catcode `\$12\catcode
  `\&12\catcode `\#12\catcode `\^12\catcode `\_12\catcode `\%12\relax}%
\providecommand \@@startlink[1]{}%
\providecommand \@@endlink[0]{}%
\providecommand \url  [0]{\begingroup\@sanitize@url \@url }%
\providecommand \@url [1]{\endgroup\@href {#1}{\urlprefix }}%
\providecommand \urlprefix  [0]{URL }%
\providecommand \Eprint [0]{\href }%
\providecommand \doibase [0]{http://dx.doi.org/}%
\providecommand \selectlanguage [0]{\@gobble}%
\providecommand \bibinfo  [0]{\@secondoftwo}%
\providecommand \bibfield  [0]{\@secondoftwo}%
\providecommand \translation [1]{[#1]}%
\providecommand \BibitemOpen [0]{}%
\providecommand \bibitemStop [0]{}%
\providecommand \bibitemNoStop [0]{.\EOS\space}%
\providecommand \EOS [0]{\spacefactor3000\relax}%
\providecommand \BibitemShut  [1]{\csname bibitem#1\endcsname}%
\let\auto@bib@innerbib\@empty
\bibitem [{\citenamefont {Suematsu}\ and\ \citenamefont {Iga}(2008)}]{1}%
  \BibitemOpen
  \bibfield  {author} {\bibinfo {author} {\bibfnamefont {Y.}~\bibnamefont
  {Suematsu}}\ and\ \bibinfo {author} {\bibfnamefont {K.}~\bibnamefont {Iga}},\
  }\href@noop {} {\bibfield  {journal} {\bibinfo  {journal} {Journal of
  Lightwave Technology}\ }\textbf {\bibinfo {volume} {26}},\ \bibinfo {pages}
  {1132} (\bibinfo {year} {2008})}\BibitemShut {NoStop}%
\bibitem [{\citenamefont {Calvez}\ \emph {et~al.}(2009)\citenamefont {Calvez},
  \citenamefont {Hastie}, \citenamefont {Guina}, \citenamefont {Okhotnikov},\
  and\ \citenamefont {Dawson}}]{2}%
  \BibitemOpen
  \bibfield  {author} {\bibinfo {author} {\bibfnamefont {S.}~\bibnamefont
  {Calvez}}, \bibinfo {author} {\bibfnamefont {J.~E.}\ \bibnamefont {Hastie}},
  \bibinfo {author} {\bibfnamefont {M.}~\bibnamefont {Guina}}, \bibinfo
  {author} {\bibfnamefont {O.~G.}\ \bibnamefont {Okhotnikov}}, \ and\ \bibinfo
  {author} {\bibfnamefont {M.~D.}\ \bibnamefont {Dawson}},\ }\href@noop {}
  {\bibfield  {journal} {\bibinfo  {journal} {Laser \& Photonics Reviews}\
  }\textbf {\bibinfo {volume} {3}},\ \bibinfo {pages} {407} (\bibinfo {year}
  {2009})}\BibitemShut {NoStop}%
\bibitem [{\citenamefont {Richter}, \citenamefont {Fried},\ and\ \citenamefont
  {Weibring}(2009)}]{3}%
  \BibitemOpen
  \bibfield  {author} {\bibinfo {author} {\bibfnamefont {D.}~\bibnamefont
  {Richter}}, \bibinfo {author} {\bibfnamefont {A.}~\bibnamefont {Fried}}, \
  and\ \bibinfo {author} {\bibfnamefont {P.}~\bibnamefont {Weibring}},\
  }\href@noop {} {\bibfield  {journal} {\bibinfo  {journal} {Laser \& Photonics
  Reviews}\ }\textbf {\bibinfo {volume} {3}},\ \bibinfo {pages} {343} (\bibinfo
  {year} {2009})}\BibitemShut {NoStop}%
\bibitem [{\citenamefont {Charlier}, \citenamefont {Blase},\ and\ \citenamefont
  {Roche}(2007)}]{4}%
  \BibitemOpen
  \bibfield  {author} {\bibinfo {author} {\bibfnamefont {J.-C.}\ \bibnamefont
  {Charlier}}, \bibinfo {author} {\bibfnamefont {X.}~\bibnamefont {Blase}}, \
  and\ \bibinfo {author} {\bibfnamefont {S.}~\bibnamefont {Roche}},\
  }\href@noop {} {\bibfield  {journal} {\bibinfo  {journal} {Reviews of Modern
  Physics}\ }\textbf {\bibinfo {volume} {79}},\ \bibinfo {pages} {677}
  (\bibinfo {year} {2007})}\ \BibitemShut {NoStop}%
\bibitem [{\citenamefont {O'Connell}\ \emph {et~al.}(2002)\citenamefont
  {O'Connell}, \citenamefont {Bachilo}, \citenamefont {Huffman}, \citenamefont
  {Moore}, \citenamefont {Strano}, \citenamefont {Haroz}, \citenamefont
  {Rialon}, \citenamefont {Boul}, \citenamefont {Noon}, \citenamefont
  {Kittrell}, \citenamefont {Ma}, \citenamefont {Hauge}, \citenamefont
  {Weisman},\ and\ \citenamefont {Smalley}}]{5}%
  \BibitemOpen
  \bibfield  {author} {\bibinfo {author} {\bibfnamefont {M.~J.}\ \bibnamefont
  {O'Connell}}, \bibinfo {author} {\bibfnamefont {S.~M.}\ \bibnamefont
  {Bachilo}}, \bibinfo {author} {\bibfnamefont {C.~B.}\ \bibnamefont
  {Huffman}}, \bibinfo {author} {\bibfnamefont {V.~C.}\ \bibnamefont {Moore}},
  \bibinfo {author} {\bibfnamefont {M.~S.}\ \bibnamefont {Strano}}, \bibinfo
  {author} {\bibfnamefont {E.~H.}\ \bibnamefont {Haroz}}, \bibinfo {author}
  {\bibfnamefont {K.~L.}\ \bibnamefont {Rialon}}, \bibinfo {author}
  {\bibfnamefont {P.~J.}\ \bibnamefont {Boul}}, \bibinfo {author}
  {\bibfnamefont {W.~H.}\ \bibnamefont {Noon}}, \bibinfo {author}
  {\bibfnamefont {C.}~\bibnamefont {Kittrell}}, \bibinfo {author}
  {\bibfnamefont {J.}~\bibnamefont {Ma}}, \bibinfo {author} {\bibfnamefont
  {R.~H.}\ \bibnamefont {Hauge}}, \bibinfo {author} {\bibfnamefont {R.~B.}\
  \bibnamefont {Weisman}}, \ and\ \bibinfo {author} {\bibfnamefont {R.~E.}\
  \bibnamefont {Smalley}},\ }\href@noop {} {\bibfield  {journal} {\bibinfo
  {journal} {Science}\ }\textbf {\bibinfo {volume} {297}},\ \bibinfo {pages}
  {593} (\bibinfo {year} {2002})}\BibitemShut {NoStop}%
\bibitem [{\citenamefont {Bachilo}\ \emph {et~al.}(2002)\citenamefont
  {Bachilo}, \citenamefont {Strano}, \citenamefont {Kittrell}, \citenamefont
  {Hauge}, \citenamefont {Smalley},\ and\ \citenamefont {Weisman}}]{6}%
  \BibitemOpen
  \bibfield  {author} {\bibinfo {author} {\bibfnamefont {S.~M.}\ \bibnamefont
  {Bachilo}}, \bibinfo {author} {\bibfnamefont {M.~S.}\ \bibnamefont {Strano}},
  \bibinfo {author} {\bibfnamefont {C.}~\bibnamefont {Kittrell}}, \bibinfo
  {author} {\bibfnamefont {R.~H.}\ \bibnamefont {Hauge}}, \bibinfo {author}
  {\bibfnamefont {R.~E.}\ \bibnamefont {Smalley}}, \ and\ \bibinfo {author}
  {\bibfnamefont {R.~B.}\ \bibnamefont {Weisman}},\ }\href@noop {} {\bibfield
  {journal} {\bibinfo  {journal} {Science}\ }\textbf {\bibinfo {volume}
  {298}},\ \bibinfo {pages} {2361} (\bibinfo {year} {2002})}\BibitemShut
  {NoStop}%
\bibitem [{\citenamefont {Mueller}\ \emph {et~al.}(2010)\citenamefont
  {Mueller}, \citenamefont {Kinoshita}, \citenamefont {Steiner}, \citenamefont
  {Perebeinos}, \citenamefont {Bol}, \citenamefont {Farmer},\ and\
  \citenamefont {Avouris}}]{7}%
  \BibitemOpen
  \bibfield  {author} {\bibinfo {author} {\bibfnamefont {T.}~\bibnamefont
  {Mueller}}, \bibinfo {author} {\bibfnamefont {M.}~\bibnamefont {Kinoshita}},
  \bibinfo {author} {\bibfnamefont {M.}~\bibnamefont {Steiner}}, \bibinfo
  {author} {\bibfnamefont {V.}~\bibnamefont {Perebeinos}}, \bibinfo {author}
  {\bibfnamefont {A.~A.}\ \bibnamefont {Bol}}, \bibinfo {author} {\bibfnamefont
  {D.~B.}\ \bibnamefont {Farmer}}, \ and\ \bibinfo {author} {\bibfnamefont
  {P.}~\bibnamefont {Avouris}},\ }\href@noop {} {\bibfield  {journal} {\bibinfo
   {journal} {Nat Nano}\ }\textbf {\bibinfo {volume} {5}},\ \bibinfo {pages}
  {27} (\bibinfo {year} {2010})}\ \BibitemShut {NoStop}%
\bibitem [{\citenamefont {Adam}\ \emph {et~al.}(2008)\citenamefont {Adam},
  \citenamefont {Aguirre}, \citenamefont {Marty}, \citenamefont {St-Antoine},
  \citenamefont {Meunier}, \citenamefont {Desjardins}, \citenamefont
  {M\'enard},\ and\ \citenamefont {Martel}}]{8}%
  \BibitemOpen
  \bibfield  {author} {\bibinfo {author} {\bibfnamefont {E.}~\bibnamefont
  {Adam}}, \bibinfo {author} {\bibfnamefont {C.~M.}\ \bibnamefont {Aguirre}},
  \bibinfo {author} {\bibfnamefont {L.}~\bibnamefont {Marty}}, \bibinfo
  {author} {\bibfnamefont {B.~C.}\ \bibnamefont {St-Antoine}}, \bibinfo
  {author} {\bibfnamefont {F.}~\bibnamefont {Meunier}}, \bibinfo {author}
  {\bibfnamefont {P.}~\bibnamefont {Desjardins}}, \bibinfo {author}
  {\bibfnamefont {D.}~\bibnamefont {M\'enard}}, \ and\ \bibinfo {author}
  {\bibfnamefont {R.}~\bibnamefont {Martel}},\ }\href@noop {} {\bibfield
  {journal} {\bibinfo  {journal} {Nano Letters}\ }\textbf {\bibinfo {volume}
  {8}},\ \bibinfo {pages} {2351} (\bibinfo {year} {2008})}\BibitemShut
  {NoStop}%
\bibitem [{\citenamefont {Xie}\ \emph {et~al.}(2009)\citenamefont {Xie},
  \citenamefont {Farhat}, \citenamefont {Son}, \citenamefont {Zhang},
  \citenamefont {Dresselhaus}, \citenamefont {Kong},\ and\ \citenamefont
  {Liu}}]{9}%
  \BibitemOpen
  \bibfield  {author} {\bibinfo {author} {\bibfnamefont {L.}~\bibnamefont
  {Xie}}, \bibinfo {author} {\bibfnamefont {H.}~\bibnamefont {Farhat}},
  \bibinfo {author} {\bibfnamefont {H.}~\bibnamefont {Son}}, \bibinfo {author}
  {\bibfnamefont {J.}~\bibnamefont {Zhang}}, \bibinfo {author} {\bibfnamefont
  {M.~S.}\ \bibnamefont {Dresselhaus}}, \bibinfo {author} {\bibfnamefont
  {J.}~\bibnamefont {Kong}}, \ and\ \bibinfo {author} {\bibfnamefont
  {Z.}~\bibnamefont {Liu}},\ }\href@noop {} {\bibfield  {journal} {\bibinfo
  {journal} {Nano Letters}\ }\textbf {\bibinfo {volume} {9}},\ \bibinfo {pages}
  {1747} (\bibinfo {year} {2009})}\BibitemShut {NoStop}%
\bibitem [{\citenamefont {Lefebvre}, \citenamefont {Homma},\ and\ \citenamefont
  {Finnie}(2003)}]{10}%
  \BibitemOpen
  \bibfield  {author} {\bibinfo {author} {\bibfnamefont {J.}~\bibnamefont
  {Lefebvre}}, \bibinfo {author} {\bibfnamefont {Y.}~\bibnamefont {Homma}}, \
  and\ \bibinfo {author} {\bibfnamefont {P.}~\bibnamefont {Finnie}},\
  }\href@noop {} {\bibfield  {journal} {\bibinfo  {journal} {Physical Review
  Letters}\ }\textbf {\bibinfo {volume} {90}},\ \bibinfo {pages} {217401}
  (\bibinfo {year} {2003})}\ \BibitemShut {NoStop}%
\bibitem [{\citenamefont {Avouris}, \citenamefont {Freitag},\ and\
  \citenamefont {Perebeinos}(2008)}]{11}%
  \BibitemOpen
  \bibfield  {author} {\bibinfo {author} {\bibfnamefont {P.}~\bibnamefont
  {Avouris}}, \bibinfo {author} {\bibfnamefont {M.}~\bibnamefont {Freitag}}, \
  and\ \bibinfo {author} {\bibfnamefont {V.}~\bibnamefont {Perebeinos}},\
  }\href@noop {} {\bibfield  {journal} {\bibinfo  {journal} {Nat Photon}\
  }\textbf {\bibinfo {volume} {2}},\ \bibinfo {pages} {341} (\bibinfo {year}
  {2008})}\ \BibitemShut {NoStop}%
\bibitem [{\citenamefont {Cahangirov}, \citenamefont {Topsakal},\ and\
  \citenamefont {Ciraci}(2010)}]{12}%
  \BibitemOpen
  \bibfield  {author} {\bibinfo {author} {\bibfnamefont {S.}~\bibnamefont
  {Cahangirov}}, \bibinfo {author} {\bibfnamefont {M.}~\bibnamefont
  {Topsakal}}, \ and\ \bibinfo {author} {\bibfnamefont {S.}~\bibnamefont
  {Ciraci}},\ }\href@noop {} {\bibfield  {journal} {\bibinfo  {journal}
  {Physical Review B}\ }\textbf {\bibinfo {volume} {82}},\ \bibinfo {pages}
  {195444} (\bibinfo {year} {2010})}\ \BibitemShut
  {NoStop}%
\bibitem [{\citenamefont {Lin}, \citenamefont {Zhuang},\ and\ \citenamefont
  {Ning}(2012)}]{13}%
  \BibitemOpen
  \bibfield  {author} {\bibinfo {author} {\bibfnamefont {Z.~Z.}\ \bibnamefont
  {Lin}}, \bibinfo {author} {\bibfnamefont {J.}~\bibnamefont {Zhuang}}, \ and\
  \bibinfo {author} {\bibfnamefont {X.~J.}\ \bibnamefont {Ning}},\ }\href@noop
  {} {\bibfield  {journal} {\bibinfo  {journal} {EPL (Europhysics Letters)}\
  }\textbf {\bibinfo {volume} {97}},\ \bibinfo {pages} {27006} (\bibinfo {year}
  {2012})}\BibitemShut {NoStop}%
\bibitem [{\citenamefont {Jin}\ \emph {et~al.}(2009)\citenamefont {Jin},
  \citenamefont {Lan}, \citenamefont {Peng}, \citenamefont {Suenaga},\ and\
  \citenamefont {Iijima}}]{14}%
  \BibitemOpen
  \bibfield  {author} {\bibinfo {author} {\bibfnamefont {C.}~\bibnamefont
  {Jin}}, \bibinfo {author} {\bibfnamefont {H.}~\bibnamefont {Lan}}, \bibinfo
  {author} {\bibfnamefont {L.}~\bibnamefont {Peng}}, \bibinfo {author}
  {\bibfnamefont {K.}~\bibnamefont {Suenaga}}, \ and\ \bibinfo {author}
  {\bibfnamefont {S.}~\bibnamefont {Iijima}},\ }\href@noop {} {\bibfield
  {journal} {\bibinfo  {journal} {Physical Review Letters}\ }\textbf {\bibinfo
  {volume} {102}},\ \bibinfo {pages} {205501} (\bibinfo {year} {2009})}\
  \BibitemShut {NoStop}%
\bibitem [{\citenamefont {Eisler}\ \emph {et~al.}(2005)\citenamefont {Eisler},
  \citenamefont {Slepkov}, \citenamefont {Elliott}, \citenamefont {Luu},
  \citenamefont {McDonald}, \citenamefont {Hegmann},\ and\ \citenamefont
  {Tykwinski}}]{15}%
  \BibitemOpen
  \bibfield  {author} {\bibinfo {author} {\bibfnamefont {S.}~\bibnamefont
  {Eisler}}, \bibinfo {author} {\bibfnamefont {A.~D.}\ \bibnamefont {Slepkov}},
  \bibinfo {author} {\bibfnamefont {E.}~\bibnamefont {Elliott}}, \bibinfo
  {author} {\bibfnamefont {T.}~\bibnamefont {Luu}}, \bibinfo {author}
  {\bibfnamefont {R.}~\bibnamefont {McDonald}}, \bibinfo {author}
  {\bibfnamefont {F.~A.}\ \bibnamefont {Hegmann}}, \ and\ \bibinfo {author}
  {\bibfnamefont {R.~R.}\ \bibnamefont {Tykwinski}},\ }\href@noop {} {\bibfield
   {journal} {\bibinfo  {journal} {Journal of the American Chemical Society}\
  }\textbf {\bibinfo {volume} {127}},\ \bibinfo {pages} {2666} (\bibinfo {year}
  {2005})}\BibitemShut {NoStop}%
\bibitem [{\citenamefont {Chalifoux}\ \emph {et~al.}(2009)\citenamefont
  {Chalifoux}, \citenamefont {McDonald}, \citenamefont {Ferguson},\ and\
  \citenamefont {Tykwinski}}]{16}%
  \BibitemOpen
  \bibfield  {author} {\bibinfo {author} {\bibfnamefont {W.~A.}\ \bibnamefont
  {Chalifoux}}, \bibinfo {author} {\bibfnamefont {R.}~\bibnamefont {McDonald}},
  \bibinfo {author} {\bibfnamefont {M.~J.}\ \bibnamefont {Ferguson}}, \ and\
  \bibinfo {author} {\bibfnamefont {R.~R.}\ \bibnamefont {Tykwinski}},\
  }\href@noop {} {\bibfield  {journal} {\bibinfo  {journal} {Angewandte Chemie
  International Edition}\ }\textbf {\bibinfo {volume} {48}},\ \bibinfo {pages}
  {7915} (\bibinfo {year} {2009})}\BibitemShut {NoStop}%
\bibitem [{\citenamefont {Chalifoux}\ and\ \citenamefont
  {Tykwinski}(2010)}]{17}%
  \BibitemOpen
  \bibfield  {author} {\bibinfo {author} {\bibfnamefont {W.~A.}\ \bibnamefont
  {Chalifoux}}\ and\ \bibinfo {author} {\bibfnamefont {R.~R.}\ \bibnamefont
  {Tykwinski}},\ }\href@noop {} {\bibfield  {journal} {\bibinfo  {journal} {Nat
  Chem}\ }\textbf {\bibinfo {volume} {2}},\ \bibinfo {pages} {967} (\bibinfo
  {year} {2010})}\ \BibitemShut {NoStop}%
\bibitem [{\citenamefont {Kertesz}, \citenamefont {Choi},\ and\ \citenamefont
  {Yang}(2005)}]{18}%
  \BibitemOpen
  \bibfield  {author} {\bibinfo {author} {\bibfnamefont {M.}~\bibnamefont
  {Kertesz}}, \bibinfo {author} {\bibfnamefont {C.~H.}\ \bibnamefont {Choi}}, \
  and\ \bibinfo {author} {\bibfnamefont {S.}~\bibnamefont {Yang}},\ }\href@noop
  {} {\bibfield  {journal} {\bibinfo  {journal} {Chemical Reviews}\ }\textbf
  {\bibinfo {volume} {105}},\ \bibinfo {pages} {3448} (\bibinfo {year}
  {2005})}\BibitemShut {NoStop}%
\bibitem [{\citenamefont {Heyd}, \citenamefont {Scuseria},\ and\ \citenamefont
  {Ernzerhof}(2003)}]{19}%
  \BibitemOpen
  \bibfield  {author} {\bibinfo {author} {\bibfnamefont {J.}~\bibnamefont
  {Heyd}}, \bibinfo {author} {\bibfnamefont {G.~E.}\ \bibnamefont {Scuseria}},
  \ and\ \bibinfo {author} {\bibfnamefont {M.}~\bibnamefont {Ernzerhof}},\
  }\href@noop {} {\bibfield  {journal} {\bibinfo  {journal} {The Journal of
  Chemical Physics}\ }\textbf {\bibinfo {volume} {118}},\ \bibinfo {pages}
  {8207} (\bibinfo {year} {2003})}\BibitemShut {NoStop}%
\bibitem [{\citenamefont {Paier}\ \emph {et~al.}(2006)\citenamefont {Paier},
  \citenamefont {Marsman}, \citenamefont {Hummer}, \citenamefont {Kresse},
  \citenamefont {Gerber},\ and\ \citenamefont {Angyan}}]{20}%
  \BibitemOpen
  \bibfield  {author} {\bibinfo {author} {\bibfnamefont {J.}~\bibnamefont
  {Paier}}, \bibinfo {author} {\bibfnamefont {M.}~\bibnamefont {Marsman}},
  \bibinfo {author} {\bibfnamefont {K.}~\bibnamefont {Hummer}}, \bibinfo
  {author} {\bibfnamefont {G.}~\bibnamefont {Kresse}}, \bibinfo {author}
  {\bibfnamefont {I.~C.}\ \bibnamefont {Gerber}}, \ and\ \bibinfo {author}
  {\bibfnamefont {J.~G.}\ \bibnamefont {Angyan}},\ }\href@noop {} {\bibfield
  {journal} {\bibinfo  {journal} {The Journal of Chemical Physics}\ }\textbf
  {\bibinfo {volume} {124}},\ \bibinfo {pages} {154709} (\bibinfo {year}
  {2006})}\BibitemShut {NoStop}%
\bibitem [{\citenamefont {Kresse}\ and\ \citenamefont
  {Furthm¨¹ller}(1996)}]{21}%
  \BibitemOpen
  \bibfield  {author} {\bibinfo {author} {\bibfnamefont {G.}~\bibnamefont
  {Kresse}}\ and\ \bibinfo {author} {\bibfnamefont {J.}~\bibnamefont
  {Furthm¨¹ller}},\ }\href@noop {} {\bibfield  {journal} {\bibinfo  {journal}
  {Physical Review B}\ }\textbf {\bibinfo {volume} {54}},\ \bibinfo {pages}
  {11169} (\bibinfo {year} {1996})}\ \BibitemShut
  {NoStop}%
\bibitem [{\citenamefont {Tani}\ \emph {et~al.}(1980)\citenamefont {Tani},
  \citenamefont {Grant}, \citenamefont {Gill}, \citenamefont {Street},\ and\
  \citenamefont {Clarke}}]{22}%
  \BibitemOpen
  \bibfield  {author} {\bibinfo {author} {\bibfnamefont {T.}~\bibnamefont
  {Tani}}, \bibinfo {author} {\bibfnamefont {P.~M.}\ \bibnamefont {Grant}},
  \bibinfo {author} {\bibfnamefont {W.~D.}\ \bibnamefont {Gill}}, \bibinfo
  {author} {\bibfnamefont {G.~B.}\ \bibnamefont {Street}}, \ and\ \bibinfo
  {author} {\bibfnamefont {T.~C.}\ \bibnamefont {Clarke}},\ }\href@noop {}
  {\bibfield  {journal} {\bibinfo  {journal} {Solid State Communications}\
  }\textbf {\bibinfo {volume} {33}},\ \bibinfo {pages} {499} (\bibinfo {year}
  {1980})}\BibitemShut {NoStop}%
\bibitem [{\citenamefont {Yannoni}\ and\ \citenamefont {Clarke}(1983)}]{23}%
  \BibitemOpen
  \bibfield  {author} {\bibinfo {author} {\bibfnamefont {C.~S.}\ \bibnamefont
  {Yannoni}}\ and\ \bibinfo {author} {\bibfnamefont {T.~C.}\ \bibnamefont
  {Clarke}},\ }\href@noop {} {\bibfield  {journal} {\bibinfo  {journal}
  {Physical Review Letters}\ }\textbf {\bibinfo {volume} {51}},\ \bibinfo
  {pages} {1191} (\bibinfo {year} {1983})}\ \BibitemShut
  {NoStop}%
\bibitem [{\citenamefont {Mikhailovskij}\ \emph {et~al.}(2013)\citenamefont
  {Mikhailovskij}, \citenamefont {Sadanov}, \citenamefont {Kotrechko},
  \citenamefont {Ksenofontov},\ and\ \citenamefont {Mazilova}}]{24}%
  \BibitemOpen
  \bibfield  {author} {\bibinfo {author} {\bibfnamefont {I.~M.}\ \bibnamefont
  {Mikhailovskij}}, \bibinfo {author} {\bibfnamefont {E.~V.}\ \bibnamefont
  {Sadanov}}, \bibinfo {author} {\bibfnamefont {S.}~\bibnamefont {Kotrechko}},
  \bibinfo {author} {\bibfnamefont {V.~A.}\ \bibnamefont {Ksenofontov}}, \ and\
  \bibinfo {author} {\bibfnamefont {T.~I.}\ \bibnamefont {Mazilova}},\
  }\href@noop {} {\bibfield  {journal} {\bibinfo  {journal} {Physical Review
  B}\ }\textbf {\bibinfo {volume} {87}},\ \bibinfo {pages} {045410} (\bibinfo
  {year} {2013})}\ \BibitemShut {NoStop}%
\bibitem [{\citenamefont {Mazilova}\ \emph {et~al.}(2010)\citenamefont
  {Mazilova}, \citenamefont {Kotrechko}, \citenamefont {Sadanov}, \citenamefont
  {Ksenofontov},\ and\ \citenamefont {Mikhailovskij}}]{25}%
  \BibitemOpen
  \bibfield  {author} {\bibinfo {author} {\bibfnamefont {T.~I.}\ \bibnamefont
  {Mazilova}}, \bibinfo {author} {\bibfnamefont {S.}~\bibnamefont {Kotrechko}},
  \bibinfo {author} {\bibfnamefont {E.~V.}\ \bibnamefont {Sadanov}}, \bibinfo
  {author} {\bibfnamefont {V.~A.}\ \bibnamefont {Ksenofontov}}, \ and\ \bibinfo
  {author} {\bibfnamefont {I.~M.}\ \bibnamefont {Mikhailovskij}},\ }\href@noop
  {} {\bibfield  {journal} {\bibinfo  {journal} {International Journal of
  Nanoscience}\ }\textbf {\bibinfo {volume} {09}},\ \bibinfo {pages} {151}
  (\bibinfo {year} {2010})}\BibitemShut {NoStop}%
\bibitem [{\citenamefont {Castelli}, \citenamefont {Salvestrini},\ and\
  \citenamefont {Manini}(2012)}]{26}%
  \BibitemOpen
  \bibfield  {author} {\bibinfo {author} {\bibfnamefont {I.~E.}\ \bibnamefont
  {Castelli}}, \bibinfo {author} {\bibfnamefont {P.}~\bibnamefont
  {Salvestrini}}, \ and\ \bibinfo {author} {\bibfnamefont {N.}~\bibnamefont
  {Manini}},\ }\href@noop {} {\bibfield  {journal} {\bibinfo  {journal}
  {Physical Review B}\ }\textbf {\bibinfo {volume} {85}},\ \bibinfo {pages}
  {214110} (\bibinfo {year} {2012})}\ \BibitemShut
  {NoStop}%
\end{thebibliography}
\end{document}